\documentclass[journal=jpcafh,manuscript=article]{achemso}
\usepackage{longtable}
\usepackage{multirow}
\usepackage{amsmath}
\usepackage{subfigure}
\usepackage[usenames,dvipsnames]{xcolor}
\usepackage{soul}
\usepackage{bm}

\usepackage{xr}
\usepackage{amssymb}
\usepackage{siunitx}
\usepackage{threeparttable}
\usepackage{multicol} \usepackage{wrapfig} \usepackage{enumitem}
\usepackage{bm} \usepackage{outlines} 

\usepackage{siunitx}\usepackage{booktabs}

\SectionNumbersOn

\author{Silvan K\"aser} \affiliation[University of Basel]{Department
  of Chemistry, University of Basel, Klingelbergstrasse 80 , CH-4056
  Basel, Switzerland.}
  \author{Markus Meuwly}
\affiliation[University of Basel]{Department of Chemistry, University
  of Basel, Klingelbergstrasse 80 , CH-4056 Basel, Switzerland.}
\email{m.meuwly@unibas.ch}

\title{Transfer-Learned Potential Energy Surfaces: Towards Microsecond-Scale Molecular Dynamics Simulations in the Gas Phase at CCSD(T) Quality}
\begin{document}
\date{\today}

\begin{abstract}
\noindent
The rise of machine learning has greatly influenced the field of
computational chemistry, and that of atomistic molecular dynamics simulations
in particular. One of its most exciting prospects is the development
of accurate, full-dimensional potential energy surfaces (PESs) for molecules and clusters,
which, however, often require thousands to tens of thousands of \textit{ab initio}
data points restricting the community to medium sized molecules and/or lower
levels of theory (e.g. DFT). Transfer learning, which improves a global PES
from a lower to a higher level of theory, offers a data efficient alternative
requiring only a fraction of the high level data (on the order of 100 are found
to be sufficient for malonaldehyde). The present work demonstrates that even with Hartree-Fock theory and a double-zeta basis set as the lower level model, transfer learning yields CCSD(T)-level quality for H-transfer barrier energies, harmonic frequencies and H-transfer tunneling splittings. Most importantly, finite-temperature molecular dynamics simulations on the sub-$\mu$s time scale in the gas phase are possible and the infrared spectra determined from the transfer learned PESs are in good agreement with experiment. It is concluded that routine, long-time atomistic simulations on PESs fulfilling CCSD(T)-standards become possible.
\end{abstract}

\section{Introduction}
Over the past decade, machine learning-based approaches have
persistently influenced the field of computational chemistry as a
whole and that of atomistic molecular dynamics (MD) simulations in
particular.\cite{keith2021combining,MM.cr:2021,unke2021machine}
Machine learning (ML) approaches begin to be routinely used in areas
ranging from reaction planning\cite{coley2018machine,chen2020retro},
to drug design\cite{blaschke2020reinvent,jimenez2020drug} and to the
construction and use of accurate local and global potential energy
surfaces
(PESs).\cite{braams2009permutationally,jiang2016potential,unke2021machine,kaser2023neural}
Common to all these applications is that they are data-driven,
i.e. they depend on the amount but also on the quality of the
available data on which the statistical models are trained. With
respect to the underlying models themselves, much flexibility exists,
encompassing kernel- or neural network-based approaches with a wide
range of architectures and specific design
choices.\cite{manzhos:2020,braams2009permutationally,jiang2016potential,MM.cr:2021,unke2021machine}\\

\noindent
The construction of a multidimensional PES suitable for
routine and robust MD simulations remains a challenging
task.\cite{vassilev2021challenges} An established approach consists of
precomputing reference energies and forces from electronic structure
methods at suitable and affordable levels of theory for a given
molecule or system, also often with particular applications such as
various spectroscopies or reaction dynamics in mind. Traditionally,
these energies were used to fit predetermined parametrized forms which
in turn not only provide the total energy of the system but also
analytical derivatives for forces required in MD simulations. The
difficulty in such approaches consists in ``guessing'' sufficiently
flexible forms of the parametrization to capture local and global
features of the PES alike.  Also, such parametrized fits are often
highly nonlinear which is another challenge, in particular for
high-dimensional PESs.\\

\noindent
ML methods, including (reproducing)
kernel\cite{ho1996general,unke2017toolkit} or neural network (NN)
approaches\cite{MM.physnet:2019,schutt2018schnet}, have been long
known to be flexible function
approximators\cite{hornik1991approximation}. It is precisely this
flexibility that is also beneficial for the representation of molecular
PESs which can have a wide range of topographies and designing
parametrizations suitable for capturing all relevant features becomes
increasingly difficult.\\

\noindent
Here, the construction, improvement and use of reactive PESs in
ring-polymer instanton (RPI) calculations and MD simulations is
discussed with the purpose of illustrating practical aspects and
probing the quality of the resulting surfaces. The method employed to
improve PESs from a lower level (LL) to a higher level (HL) of theory
is transfer learning
(TL)\cite{taylor2009transfer,smith2019approaching,pan2009survey} which
is rooted in the notion that the overall topographies of molecular
PESs at different levels of theory are often similar. As an example,
it is often found that the number of local minima is comparable or
even identical if quantum chemical methods at different levels of
theory are employed, whereas the geometrical structures of course
differ. This topographical similarity can be exploited in TL
approaches in that measured amounts of HL information are used to
locally distort the model based on LL data. One particular focus in
the present work is on studying the influence of the level of
electronic structure theory used to construct the surrogate LL model
on the quality of the HL PES and making best use of the data
generated. To the best of our knowledge this contribution presents the
first systematic analysis of the influence of the LL PES in TL of
global reactive molecular potential energy surfaces.\\

\noindent
Malonaldehyde (MA) is a well-studied molecule
experimentally
\cite{firth1991tunable,baba1999detection,baughcum1984microwave,turner1984microwave,smith1983infrared,firth1989matrix,chiavassa1992experimental,duan2004high}
and from
computations\cite{wang2008full,MM.ma:2010,mizukami2014compact,MM.ma:2014,mm.ht:2020} and provides a
suitable benchmark system for the present work and recent
results\cite{mm.tlrpimalonaldehyde:2022} serve as a reference. The
properties used for
evaluating the quality of the final PESs include energies
(e.g. RMSE($E$) from comparison to \textit{ab initio} reference),
barrier heights for hydrogen transfer (H-transfer), harmonic frequencies $\omega$
(allowing comparison with
experiments\cite{smith1983infrared,luttschwager2013vibrational} and
\textit{ab initio} results\cite{mm.tlrpimalonaldehyde:2022}), and
tunneling splittings $\Delta_{\rm H/D}$ (allowing comparison to
experiment\cite{firth1991tunable,baba1999detection,baughcum1984microwave}
and
literature\cite{mm.tlrpimalonaldehyde:2022}). Alternative/additional
observables include H-transfer rates and infrared spectra inferred
from MD simulations which can also be compared with experiment.\\

\noindent
The present work is structured as follows. First, the methods
including the NN-based representation of the PESs and TL are
discussed. Then, the results comprising out of sample errors, energy
barriers, harmonic frequencies and tunneling splittings as well as the
results from an aggregate of 1.5~$\mu$s of MD simulations are
presented. Finally, the results are discussed in a broader context and
compared to previous experiments and computational work.\\

\section{Methods}
This section introduces the employed ML approaches, the data
generation procedure, RPI theory, the MD simulation protocol, analysis
and computation of the observables.

\subsection{Machine Learning}
All full-dimensional, reactive PESs used in this work are represented
by a high dimensional, message-passing\cite{gilmer2017neural} NN of the PhysNet type\cite{MM.physnet:2019}. PhysNet predicts
energies, forces and dipole moments for arbitrary molecular
configurations from a feature vector that is learned to describe the
local chemical environment (here, up to a cutoff radius $r_{\rm cut} =
10$~\r{A}) of each atom $i$. The total potential energy of a molecule
with $N$ atoms for a given geometry is given by
\begin{align}
    E = \sum_{i=1}^N E_i + k_e \sum_{i=1}^N\sum_{j>i}^N\frac{q_iq_j}{r_{ij}}
\end{align}
where $E_i$ are the atomic contributions to the total energy of the
molecule, $k_e$ is Coulomb's constant and the second term describes
pairwise electrostatic contributions. Note that the partial charges
$q_i$ are corrected for charge conservation and that the electrostatic
contribution is damped at small inter-atomic distances $r_{ij}$ due to
the singularity (for details see
Reference\citenum{MM.physnet:2019}). The forces $\bm{F}$ and Hessians
$\bm{H}$ are obtained analytically from reverse mode automatic
differentiation\cite{baydin2017automatic} as implemented in
Tensorflow\cite{abadi2016tensorflow} and the dipole moment is
calculated as
\begin{align}
    \bm{\mu} = \sum_{i=1}^N q_i\bm{r_i}. 
\end{align}

\noindent
The learnable parameters in PhysNet were fit to reference \textit{ab
  initio} energies, forces and dipole moments following the procedure
described in Reference~\citenum{MM.physnet:2019}. Note that the
partial charges $q_i$ were adapted to best reproduce the dipole moment
from the \textit{ab initio} calculations.\\

\noindent
Recently, TL\cite{pan2009survey,tan2018survey} and
related approaches including $\Delta$-ML\cite{fu:2008,DeltaPaper2015,bowman2022deltaperspective}
have become very popular and successful in improving a given LL PES to
a HL of theory. At its core, TL builds on the knowledge acquired by
solving one task (representing the LL PES) to solve a new, related
task (representing the HL PES).\cite{pan2009survey} In practice, the
parameters of the LL PES are used as a good initial guess and are
fine-tuned with little, though judiciously, chosen HL information.  In
this work, all parameters of PhysNet were allowed to change in the TL
step.

\subsection{\textit{Ab Initio} Data}
All molecular structures that were used to in the fitting procedure of PhysNet
were available from earlier work\cite{mm.ht:2020} and their sampling
process is described in detail in Reference~\citenum{mm.ht:2020}. The
data set contains structures for MA, acetoacetaldehyde (3-oxobutanal),
acetylacetone (pentan-2, 4-dion) as well as a total of 49
substructures as motivated by the "amon"
approach.\cite{huang2020quantum} An initial set of structures was
generated by running Langevin dynamics at 1000~K using the
semi-empirical PM7 method\cite{stewart2007optimization} as implemented
in MOPAC\cite{stewar2016mopac}. Next, \textit{ab initio} energies,
forces and dipole moments were determined at the MP2/aug-cc-pVTZ
(aVTZ) level of theory using MOLPRO\cite{MOLPRO} and were used to
train two preliminary PhysNet models.  The preliminary models were
then used to suitably augment the data set using adaptive
sampling\cite{csanyi2004learn}. The final data set contained MP2/aVTZ
level energies, forces and molecular dipole moments for a total of
71\,208 structures including all three molecules and their amons. This
PES is referred to as PhysNet$^{\rm MP2}$.\\

\noindent
To be able to study the influence of the LL on the quality of the HL,
the data set containing 71\,208 structures from earlier
work\cite{mm.ht:2020} was recalculated at two additional "low" levels
of theory, namely HF/cc-pVDZ (VDZ) and B3LYP/aVTZ using
MOLPRO\cite{MOLPRO}. The two new data sets are used to learn
representations of the PES using PhysNet and are termed PhysNet$^{\rm
  HF}$ and PhysNet$^{\rm B3LYP}$, respectively. The relative CPU times
for the \textit{ab initio} calculations (that include energies, forces
and dipole moments) are roughly 1:10:170:4000 for the HF/VDZ,
B3LYP/aVTZ, MP2/aVTZ and CCSD(T)/aVTZ levels of theory,
respectively.\\

\noindent
The CCSD(T)/aVTZ level of theory was chosen as higher level of theory,
which is computationally much more demanding, and will serve as target
in the TL step.\cite{mm.tlrpimalonaldehyde:2022} Although the
determination of forces and dipole moments is roughly an order of
magnitude more expensive than an energy-only calculation, the
information gained from including gradients (27 components for
malonaldehyde) and dipole moments is considerable. The data set sizes
employed in the TL step of the present work include $N = [25, 50, 100,
  862]$ (corresponding to TL$_0$, TL$_1$, TL$_2$ and TL$_{\rm ext}$)
MA structures. Except for $N=862$ (which corresponds to an extended
data set that contains all HL
information.\cite{mm.tlrpimalonaldehyde:2022}) the MA structures were
augmented iteratively and are specifically chosen to ideally
cover all relevant spatial
regions (i.e. structures on and around the minimum energy path, instanton
path, global minimum and transition state)
of the PESs that were found to be important for the
determination of the tunneling
splitting\cite{mm.tlrpimalonaldehyde:2022}.  The resulting TL PESs
corresponding to the LL models (PhysNet$^{\rm HF}$, PhysNet$^{\rm
  B3LYP}$ or PhysNet$^{\rm MP2}$) are termed TL$^{\rm x}_{\rm y}$ with
x$\in$[HF, B3LYP, MP2] and y$\in$[0, 1, 2, ext].\\

\subsection{Ring-Polymer Instanton Theory}
The ring polymer instanton (RPI) method offers a semiclassical
approximation for computing tunneling splittings in molecular
systems.\cite{tunnel,InstReview,hexamerprism,richardson2017full}
Instanton theory, which is closely related to the WKB
approximation\cite{Milnikov2001} in a one-dimensional model can be
applied to multidimensional systems as well. It identifies the optimal
tunneling pathway, called the instanton, by locating the
imaginary-time $\tau \rightarrow \infty$ path connecting two
degenerate wells that minimizes the action, $S$. To
determine the instanton, a ring-polymer optimization technique is used
to discretize the path into $N$ beads and take the limit as
$N\rightarrow\infty$. $S$ is then determined using information from
the potential energy of and distances between neighboring beads along
the instanton path (IP). Usually, the instanton path does not pass
through the saddle point of the reaction and is different from the
minimum energy path. In addition to the IP, a contribution $\Phi$
characterizing the fluctuations around the path needs to be computed to
second order which necessitates the availability of Hessians at each
bead. The tunneling splitting is then given by
\begin{align}
    \Delta = \frac{2\hbar}{\Phi}\sqrt{\frac{S}{2\pi\hbar}}e^{-S/\hbar}.
\end{align}
Full technical details regarding the RPI approach are given, e.g., in
References\citenum{richardson2011ring} and
\citenum{richardson2018ring}.\\

\subsection{Molecular Dynamics Simulations}
The MD simulations evaluated in this work were all performed using
PhysNet PESs and using the atomic simulation environment
(ASE\cite{larsen2017atomic}) in Python. Starting from a given PES, the
structure of MA was first optimized to its minimum energy structure
before random momenta were drawn from a Maxwell-Boltzmann distribution
corresponding to $T = [300, 500]$~K and assigned to the atoms. The MD
simulations were carried out in the \textit{NVE} ensemble using the
Velocity Verlet algorithm and a time step of $\Delta t = 0.25$~fs to
conserve energy as bonds involving hydrogen were flexible. The system
was first allowed to equilibrate for 2.5~ps and was followed by a
production simulation of 250~ps each. At each temperature and using a
representative HL PES for each class of TL PESs, 1000 independent
trajectories were run using different initial momenta. This
accumulated to a total of 750~ns simulation time per temperature
$T$.\\

\noindent
The MD simulations were used to estimate H-transfer rates,
which were calculated as the accumulated number of H-transfers divided by the
accumulated total simulation time, $N_{\rm HT} / t_{\rm tot}$.  A H-transfer
was considered complete whenever the H-atom transfers from being
closer to one O-Atom to the other (e.g. when $r_{\rm O_A H} < r_{\rm
  O_B H}$ changes to $r_{\rm O_A H} > r_{\rm O_B H}$), although other
criteria exist and have been used.\cite{MM.ammonia:2002} For the
present work the details of the criterion are of less interest because
only the relative rates from simulation based on the different PESs is
analyzed. The same trajectories were also used to compute infrared
(IR) spectra from the Fourier transform of the dipole-dipole
auto-correlation function,\cite{topfer2023molecular} according to
\begin{equation}
  I(\omega) n(\omega) \propto Q(\omega) \cdot \mathrm{Im}\int_0^\infty
  dt\, e^{i\omega t} 
  \sum_{i=x,y,z} \left \langle \boldsymbol{\mu}_{i}(t)
  \cdot {\boldsymbol{\mu}_{i}}(0) \right \rangle.
\label{eq:IR}
\end{equation}
Here, the Fourier transform was further corrected by a quantum
correction factor\cite{kumar2004quantum} $Q(\omega) = \tanh{(\beta
  \hbar \omega / 2)}$.\\

\section{Results}
The following section presents first the results derived from the LL
PESs for MA. This includes out of sample errors, energy barriers and
harmonic frequencies as determined from the LL PESs including the
comparison to their \textit{ab initio} references, and tunneling
splittings. This is followed by the results for the HL PESs.

\begin{figure}[h!]
\centering
\includegraphics[width=0.7\textwidth]{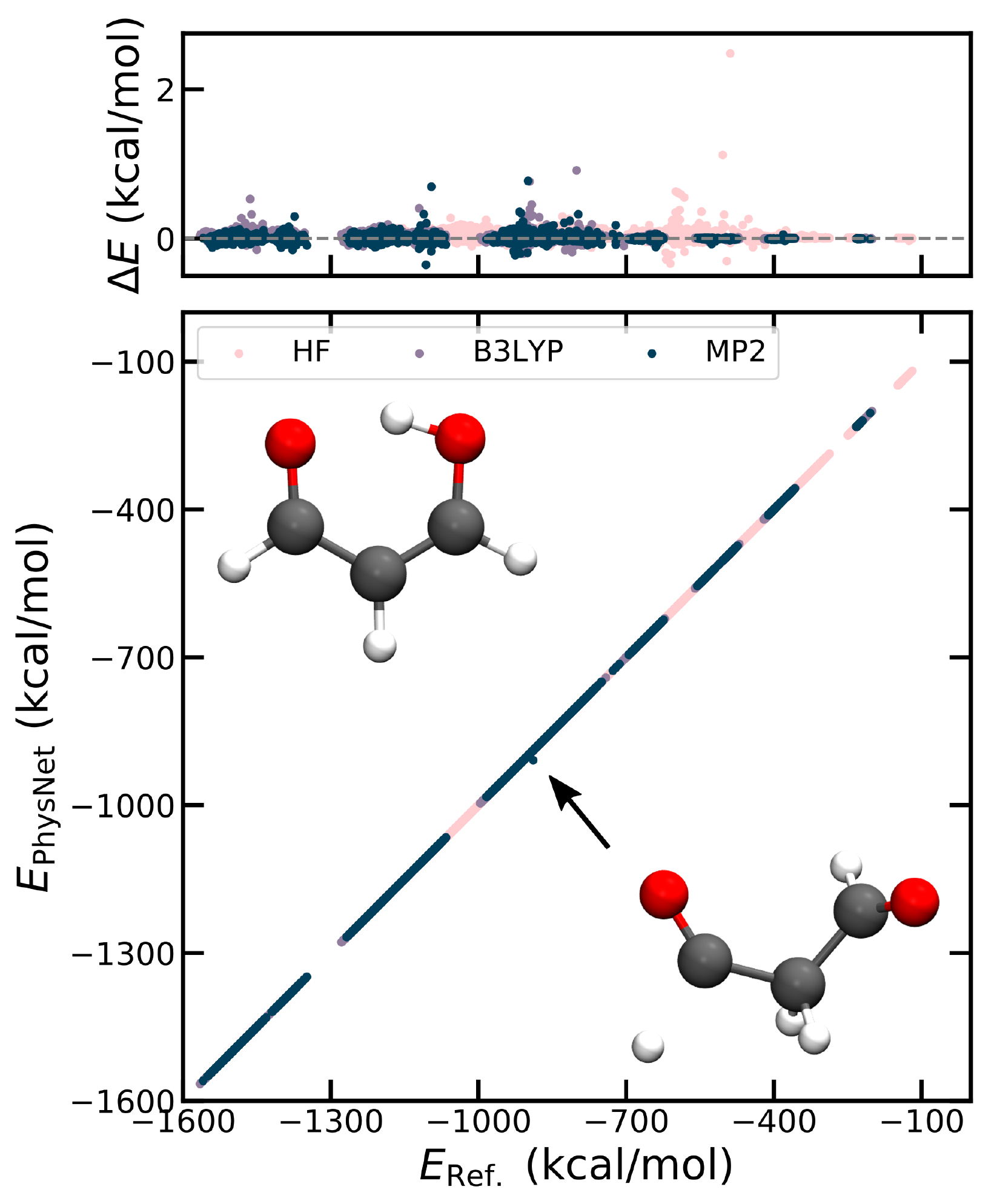}
\caption{The main panel shows the out of sample
  performance of the LL models. PhysNet models are trained on
  energies, forces and dipole moments determined at the HF/VDZ, B3LYP/aVTZ and MP2/aVTZ level of
  theory for a total 71\,208 structures. 9208 structures ($\sim
  13~\%$) served as test set and were not used during training. The
  test set contains structures for MA, acetoacetaldehyde,
  acetylacetone and the amons. The top panel reports $\Delta = E_{\rm
    Ref.} - E_{\rm PhysNet}$ and contains all but a single outlier
  from the MP2 model (that has $\Delta\approx20$~kcal/mol) and is
  shown in the lower right corner of the main panel at $\sim -800$~kcal/mol) which was omitted
  for better visibility of the errors. The molecular structure in the upper left corner
  shows the global minimum of MA.}
\label{fig:errcorr_LL_models}
\end{figure}

\subsection{Quality of LL PESs}
The LL PESs generated at three different levels of theory including
HF, B3LYP and MP2 (PhysNet$^{\rm HF}$, PhysNet$^{\rm B3LYP}$ and
PhysNet$^{\rm MP2}$) are evaluated on a test set containing 9208
structures. The mean absolute (MAE) and root mean squared (RMSE)
errors on energies and forces are summarized in
Table~\ref{tab:outofsample_errors}. MAE($E$) for all models are below
0.05~kcal/mol and are lowest for PhysNet$^{\rm MP2}$ (0.02~kcal/mol).
Conversely, the RMSE($E$) for PhysNet$^{\rm MP2}$ is highest (a factor
of $\sim 5$ larger than for PhysNet$^{\rm HF}$ and PhysNet$^{\rm
  B3LYP}$). The MAE($F$) are within $0.008$~kcal/mol/\r{A} for all
three LL PESs while the RMSE($F$) differ by $\sim 0.1$~kcal/mol/\r{A}
at most. The correlation and the errors of the PhysNet energies with
respect to the respective reference \textit{ab initio} values are
shown in Figure~\ref{fig:errcorr_LL_models} for the test set.\\

\noindent
The barrier for H-transfer from the {\it ab initio} calculations varied
between 9.96, to 2.98 and 2.74~kcal/mol for the HF, B3LYP and MP2
level of theory, respectively. Excellent fitting accuracy with respect
to \textit{ab initio} reference is found for all LL PESs with the
largest absolute deviation of $0.05$~kcal/mol for PhysNet$^{\rm
  MP2}$. These barrier heights compare with 4.03 kcal/mol and 4.09
kcal/mol from high-level CCSD(T) calculations of different
flavours.\cite{braams2009permutationally,mizukami2014compact}\\

\begin{figure}[h!]
\centering
\includegraphics[width=1.0\textwidth]{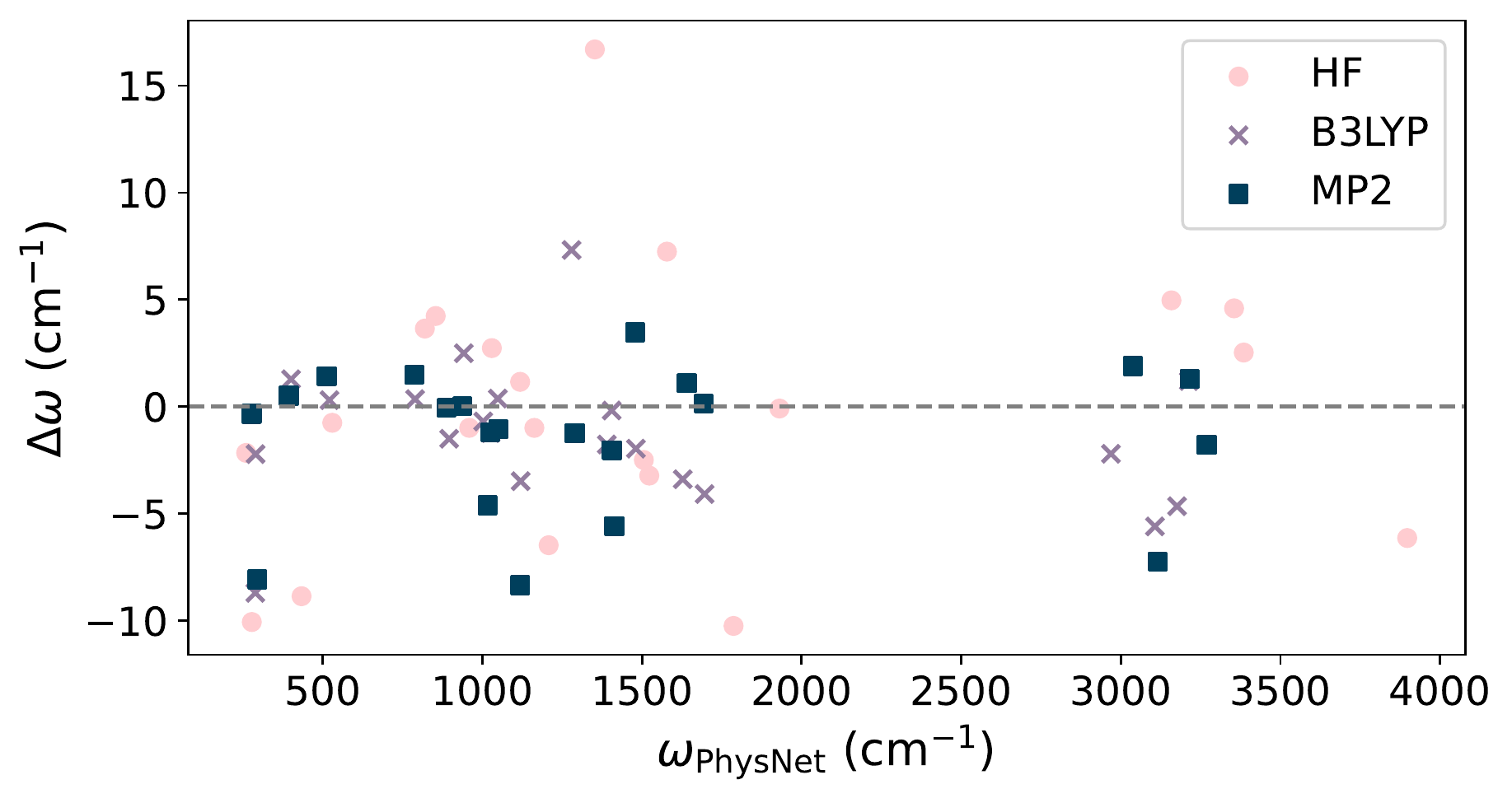}
\caption{Harmonic frequencies $\omega$ of MA together with their
  representation errors $\Delta\omega = \omega_{\rm Ref.} -
  \omega_{\rm PhysNet}$ determined from PhysNet trained on reference
  data at the LL methods. A corresponding plot for the frequencies of
  the H-transfer transition state is given in
  Figure~\ref{sifig:err_harmonic_frequencies_ts} and all harmonic
  frequencies are listed in
  Tables~\ref{sitab:harmonic_frequencies_opt} and
  \ref{sitab:harmonic_frequencies_TS}.}
\label{fig:err_harmonic_frequencies_min}
\end{figure}

\noindent
Harmonic frequencies constitute an additional measure for the accuracy
of a PES around a stationary point. The MAEs and the RMSEs for the
harmonic frequencies of the global minimum of MA for all LL PESs are
reported in Table~\ref{tab:outofsample_errors}. While the MAEs (RMSEs)
for all LL PESs are below 5 (7)~cm$^{-1}$, the errors for
PhysNet$^{\rm HF}$ are slightly larger by approximately a factor of 2
compared to PhysNet$^{\rm B3LYP}$ and PhysNet$^{\rm MP2}$, which are
within 0.1~cm$^{-1}$ of one another. The actual deviation of the
PhysNet harmonic frequencies with respect to their reference is shown
in Figure~\ref{fig:err_harmonic_frequencies_min} for the global
minimum of MA. The largest difference is found for PhysNet$^{\rm HF}$
with a deviation of $\sim 17$~cm$^{-1}$. For the H-transfer transition state,
the MAEs (RMSEs) are all below 6 (9)~cm$^{-1}$ (see
Table~\ref{sitab:harmonic_frequencies_TS}). Here, PhysNet$^{\rm
  B3LYP}$ yields the most accurate frequencies with a MAE and RMSE of
2.3 and 3.6~cm$^{-1}$, respectively.  A comprehensive list of the
harmonic frequencies for the minimum and H-transfer transition state as
calculated on the LL PhysNet PESs and their \textit{ab initio}
reference are given in Tables~\ref{sitab:harmonic_frequencies_opt} and
\ref{sitab:harmonic_frequencies_TS}, respectively.  \\

\noindent
As one of the target observables, that allows a comparison to
experiment\cite{firth1991tunable,baba1999detection,baughcum1984microwave},
the tunneling splittings for H-transfer and deuterium transfer (D-transfer)
$\Delta_{\rm H/D}$ are calculated for all LL PESs. The tunneling
splitting is exquisitely sensitive to the quality and accuracy of the
PES along and around the instanton path comprising the minimum energy
structure and the region near the transition state. The
tunneling splitting for H-transfer, $\Delta_{\rm H}$, ranges from $0.036$, to
71.1 and to 96.3~cm$^{-1}$ for PhysNet$^{\rm HF}$, PhysNet$^{\rm
  B3LYP}$ and PhysNet$^{\rm MP2}$. Similarly, the tunneling splittings
for D-transfer, $\Delta_{\rm D}$, are 9.9$\cdot10^{-4}$ , 12.8 and
18.8~cm$^{-1}$. This compares with
experimentally\cite{firth1991tunable,baba1999detection} determined
tunneling splittings of 21.583 and 2.915~cm$^{-1}$ for H-transfer and D-transfer,
respectively, and RPI calculations at CCSD(T) quality (TL +
RPI)\cite{mm.tlrpimalonaldehyde:2022} of 25.3/3.7~cm$^{-1}$. The
hydrogen tunneling splitting $\Delta_{\rm H}$ spans more than three
orders of magnitude and impressively illustrates its exquisite
sensitivity to the shape and accuracy of the PES.\\

\begin{table}[h!]
\begin{tabular}{c|rrr}\toprule
      &  \textbf{HF}   &  \textbf{B3LYP} & \textbf{MP2}\\\midrule
MAE($E$)/(kcal/mol)      &  0.046 & 0.028 & 0.020 \\
RMSE($E$)/(kcal/mol)      &  0.063 & 0.040 & 0.210\\
MAE($F$)/(kcal/mol/\AA\/)       &  0.067 & 0.059 & 0.062\\
RMSE($F$)/(kcal/mol/\AA\/)      &  0.177 & 0.141 & 0.242\\\midrule
$E_{\rm B}$/(kcal/mol)             & 10.00 & 3.02 & 2.79\\
$E_{\rm B}^{\rm Ref.}$/(kcal/mol)  &  9.96 & 2.98 & 2.74\\\midrule
MAE($\omega$)/(cm$^{-1}$)      &  4.8 & 2.6 & 2.5 \\
RMSE($\omega$)/(cm$^{-1}$)     &  6.2 & 3.5 & 3.6\\\midrule
$\Delta_{\rm H}$/(cm$^{-1}$) & 0.0358 & 71.1 & 96.3\\
$\Delta_{\rm D}$/(cm$^{-1}$) & 0.000991 & 12.8 & 18.8 \\\bottomrule
\end{tabular}
\caption{Summary of the performance of the LL PhysNet PESs: Out of
  sample errors for predicting the test set of 9208 structures for
  which energies and forces have been determined at the three
  different levels of theory. Energy barrier $E_{\rm B}$ for H-transfer
  obtained from PhysNet and from \textit{ab initio} calculations
  $E_{\rm B}^{\rm Ref.}$. MAEs and RMSEs for harmonic frequencies
  obtained from the LL PES with respect to appropriate reference
  \textit{ab initio} frequencies. 
  Tunneling splittings for H-transfer and D-transfer obtained from PhysNet $\Delta$
  are listed. The
  experimentally\cite{firth1991tunable,baba1999detection} determined
  tunneling splitting is 21.583 and 2.915~cm$^{-1}$ for H-transfer and
  D-transfer, respectively, which compares to CCSD(T) quality
  predictions (TL + RPI)\cite{mm.tlrpimalonaldehyde:2022} of
  25.3/3.7~cm$^{-1}$.}\label{tab:outofsample_errors}
\end{table}

\subsection{Performance of the TL PESs}
\begin{figure}[h!]
\centering
\includegraphics[width=1.0\textwidth]{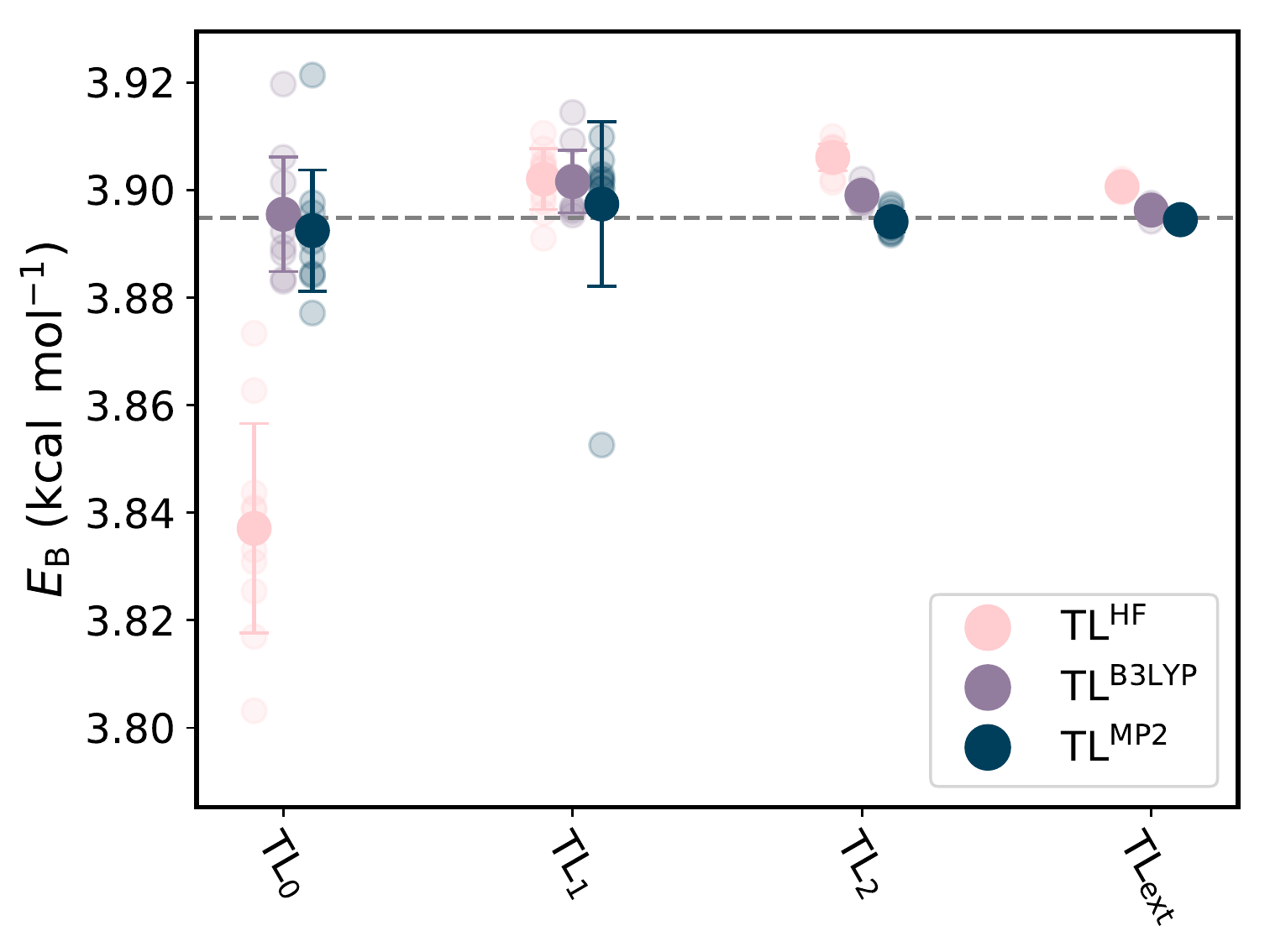}
\caption{Energy barriers $E_{\rm B}$ for H-transfer from all TL
  PESs (transparent circles). The corresponding averages (opaque
  circle) and standard deviations (error bars as $\pm\sigma$) are
  indicated as well.  The different colors represent TLs that were
  started based on LL PESs of different levels of theory, including
  HF, B3LYP and MP2. The gray dashed line corresponds to the
  \textit{ab initio} barrier height for H-transfer at the CCSD(T)/aug-cc-pVTZ
  level of theory and is $E_{\rm B} = 3.8948$~kcal/mol. The energy
  barriers are accurate already for TL$_0$ (25 data points), except
  for TL$^{\rm HF}$. Presumably, this is caused by the rather big
  difference between the HF and the (target) CCSD(T) energy barrier.}
\label{fig:TL_energy_barriers}
\end{figure}

The quality of the HL, transfer-learned PESs TL$^{\rm x}_{\rm y}$ with
x$\in$[HF, B3LYP, MP2] and y$\in$[0, 1, 2, ext] starting form three
different LL PESs is broadly evaluated next. On account of the small
HL data set sizes (including $N = [25, 50, 100, 862]$ for TL$_0$,
TL$_1$, TL$_2$ and TL$_{\rm ext}$), each TL is repeated 10 times on
different splits of the data. The energy barriers $E_{\rm B}$ for the
different HL PESs together with their variation are shown in
Figure~\ref{fig:TL_energy_barriers} for all TLs (transparent
circles). Their averages and standard deviations are indicated as
opaque circles and error bars, respectively. The gray dashed line
corresponds to the \textit{ab initio} CCSD(T)/aVTZ energy barrier
determined in previous work\cite{mm.tlrpimalonaldehyde:2022}. While
TL$^{\rm B3LYP}_0$ and TL$^{\rm MP2}_0$ (i.e. requiring only 25 data
points) already yield $E_B$ within 0.003~kcal/mol of the \textit{ab
  initio} CCSD(T)/aVTZ barrier, the deviation of $\sim 0.06$~kcal/mol
for TL$^{\rm HF}_0$ is slightly larger. For all larger data set sizes,
the energy barrier for H-transfer of the HL PESs are accurate and equal among
themselves to within $\sim 0.01$~kcal/mol. Overall, TL from all LL
models yields HL models with $E_{\rm B}$ for H-transfer to within
$0.06$~kcal/mol of the reference value from \textit{ab initio}
CCSD(T)/aVTZ calculations determined in previous
work\cite{mm.tlrpimalonaldehyde:2022}. This \textit{ab initio}
CCSD(T)/aVTZ barrier of 3.895~kcal/mol compares to values of 4.09 and
4.03~kcal/mol as obtained at the CCSD(T)/aug-cc-pV5Z (optimization
carried out at CCSD(T)/aVTZ level) and fc-CCSD(T) (F12*)/def2-TZVPP
level of theory,
respectively.\cite{braams2009permutationally,mizukami2014compact}\\

\begin{figure}[h!]
\centering
\includegraphics[width=1.0\textwidth]{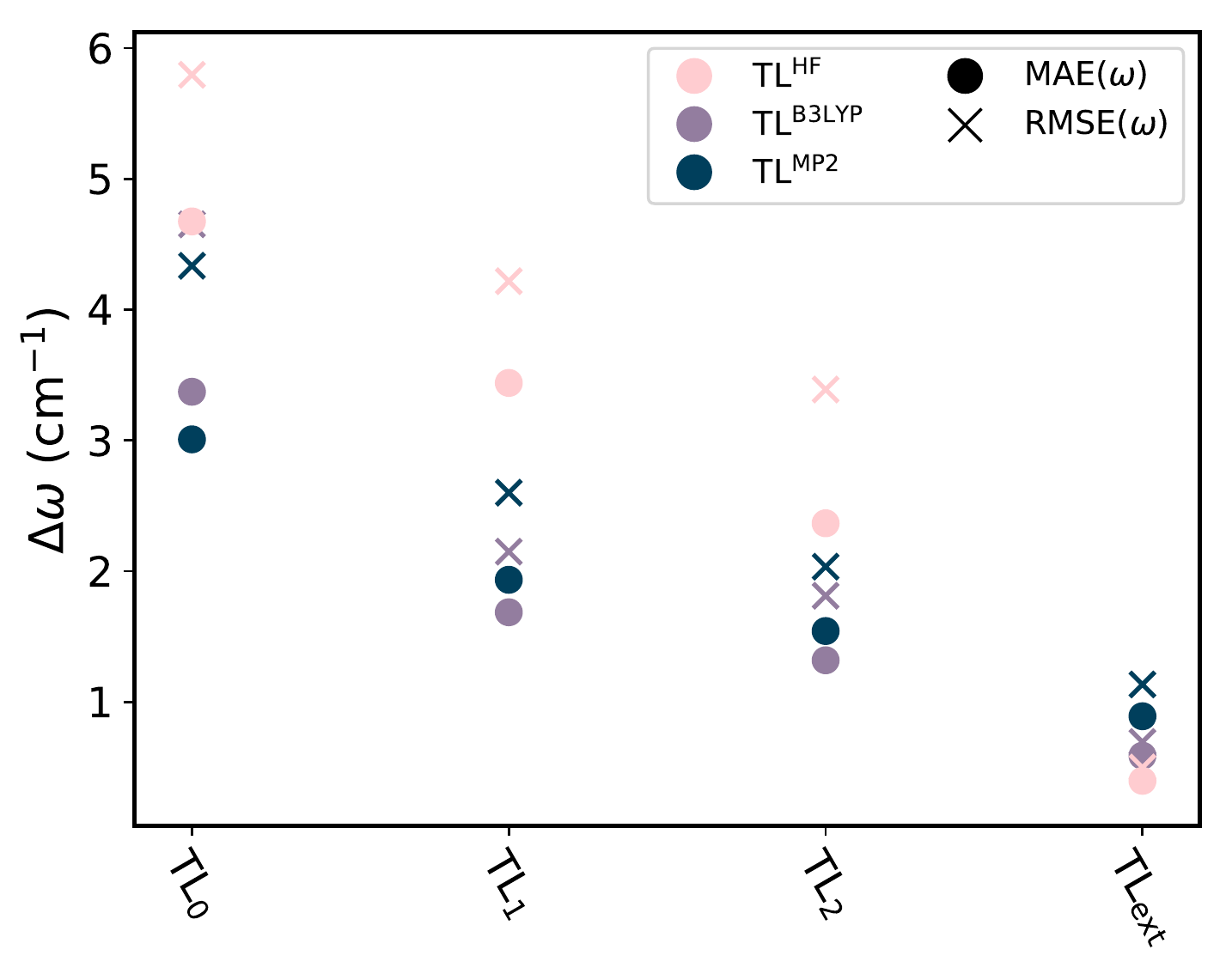}
\caption{The ensemble prediction of harmonic frequencies for the
  optimized structure of MA obtained from the different TL$_{\rm y}$
  (y$\in$[0,1,2,ext]) models are compared to the reference harmonic
  frequencies calculated at the CCSD(T)/aVTZ level of theory. Mean
  absolute errors (MAE) and root mean squared errors (RMSE) are
  reported. The "null hypothesis" (i.e. the RMSE of the \textit{ab
    initio} HF/B3LYP/MP2 harmonic frequencies with respect to the
  CCSD(T) frequencies) yields an RMSE of 177, 28 and 21~cm$^{-1}$ for
  HF, B3LYP and MP2, respectively.}
\label{fig:TL_mae_rmse_harmfreq}
\end{figure}

\noindent
The energy barrier $E_{\rm B}$ for H-transfer is a local, rather low
dimensional property of a PES. Conversely, harmonic frequencies for
stationary points probe the curvature of the PES in all
directions. Figure~\ref{fig:TL_mae_rmse_harmfreq} shows the MAE and
RMSEs of ensemble predictions (i.e. an average over 10 TLs) for the
global minimum of MA of TL$^{\rm HF}$, TL$^{\rm B3LYP}$ and TL$^{\rm
  MP2}$ for the different TL data set sizes (i.e. 25, 50, 100, 862).
Averaged errors below 6~cm$^{-1}$ are achieved for all HL PESs with as
little as 25 data points. TL with 100 CCSD(T)/aVTZ data
points the MAEs (RMSEs) are further reduced to 2.4, 1.3 and 1.5 (3.4,
1.8 and 2.0)~cm$^{-1}$ for TL$^{\rm HF}_2$, TL$^{\rm B3LYP}_2$ and
TL$^{\rm MP2}_2$, respectively. The RMSEs between the \textit{ab
  initio} HF/B3LYP/MP2 harmonic frequencies and the high-level CCSD(T)
frequencies are 177, 28 and 21~cm$^{-1}$ for HF, B3LYP and MP2,
respectively, and provides an impression by how much the shape of the
LL PESs need to adjust around the minimum at the TL
step.\\

\begin{figure}[h!]
\centering
\includegraphics[width=1.0\textwidth]{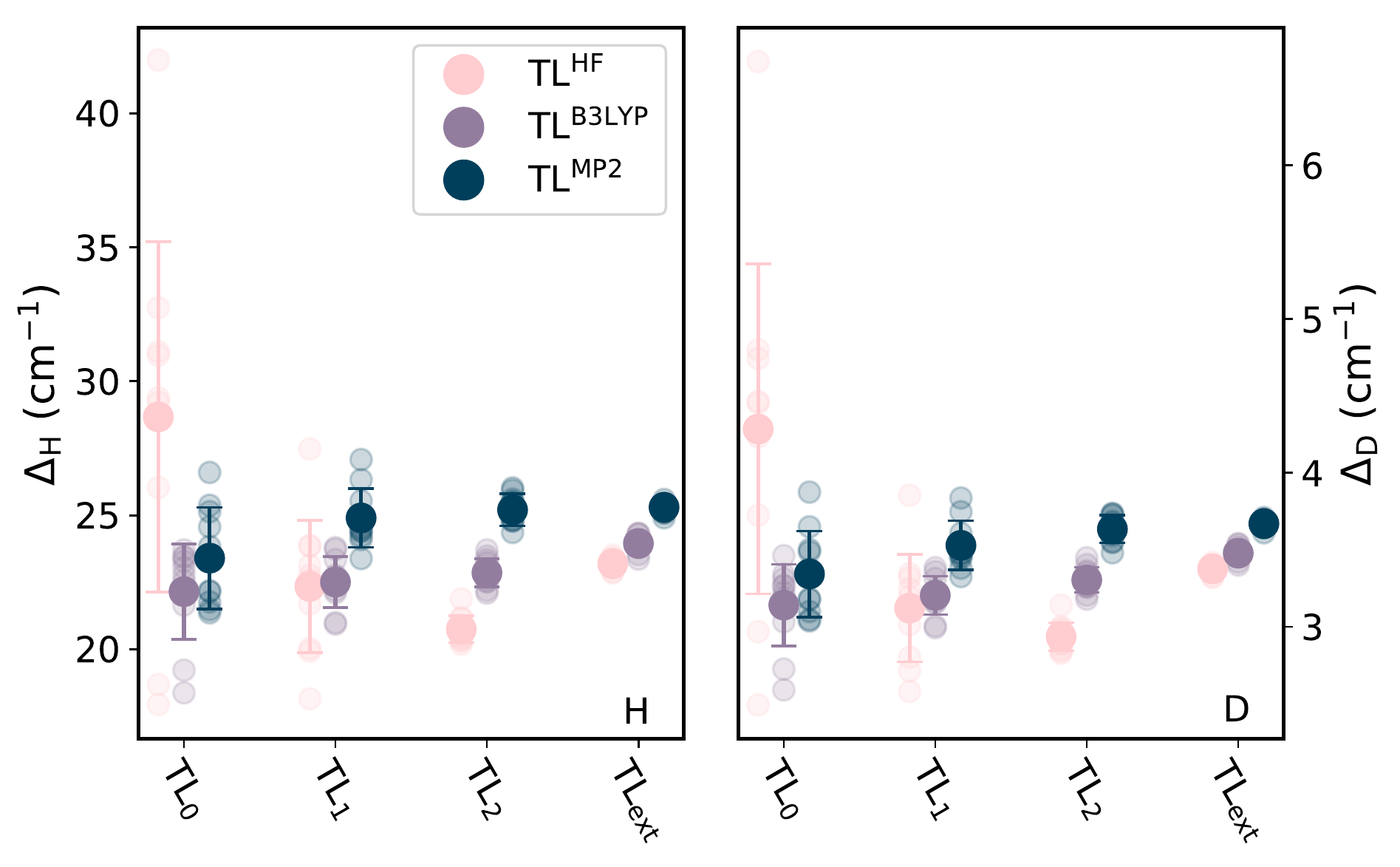}
\caption{Tunneling splittings for H- and D-transfer (left and right
  panels) from all TL PESs (transparent circles).  The corresponding
  averages (opaque circle) and standard deviations (error bars as
  $\pm\sigma$) are indicated as well.  The different colors represent
  TLs that were started based on LL PESs of different levels of
  theory, including HF, B3LYP and MP2.}
\label{fig:TL_splittings}
\end{figure}

\noindent
The tunneling splitting is sensitive to the local dynamics and the
PES's topology around the H-transfer instanton path. Hydrogen and deuterium
tunneling splittings were used as target observables in recent work
for the iterative and systematic construction of different data sets
based on PhysNet$^{\rm MP2}$. \cite{mm.tlrpimalonaldehyde:2022}
Figure~\ref{fig:TL_splittings} presents the tunneling splittings for
H-transfer and D-transfer as calculated from all different HL PESs (TL$^{\rm HF}_0$ to
TL$^{\rm MP2}_{\rm ext}$).  Again, their averages and standard
deviations are shown as opaque circles and error bars,
respectively. As had been found for $E_{\rm B}$, the fluctuation is
largest for TL$^{\rm HF}_0$ with tunneling splittings ranging from
$\sim 18$ to 42~cm$^{-1}$. This fluctuation is smaller by a factor of
$\sim 5$ for both TL$^{\rm B3LYP}_0$ and TL$^{\rm MP2}_0$. For TL$_2$
the fluctuation among the three classes of HL PES become
comparable. Average $\Delta_{\rm H}$ of 20.7, 22.9 and 25.2~cm$^{-1}$
are found for TL$^{\rm HF}_2$, TL$^{\rm B3LYP}_2$ and TL$^{\rm
  MP2}_2$. Increasing the data set size to 862 (TL$_{\rm ext}$)
further reduces both the differences within the classes and the
standard deviation of the ensemble predictions. Here, average
$\Delta_{\rm H}$ of 23.2, 23.9 and 25.3~cm$^{-1}$ are found for
TL$^{\rm HF}_{\rm ext}$, TL$^{\rm B3LYP}_{\rm ext}$ and TL$^{\rm
  MP2}_{\rm ext}$, respectively. It is noted, that $\Delta_{\rm H}$
from TL$^{\rm MP2}$ only marginally changed upon increasing the TL
data set size from 100 to 862 (TL$_2$ to TL$_{\rm ext}$). Similar
observations hold for D-transfer, which indicates that TL$^{\rm MP2}_2$ is
basically converged with MP2 as the LL model.\\

\subsection{Molecular Dynamics on the HL PESs}
Further evaluations focus on dynamical properties derived from MD
simulations carried out on the HL PESs. As a TL data set size of 100
CCSD(T) data points (TL$_2$) is a very realistic scenario for final
predictions (i.e. allows to check for convergence of the results and a
small enough data set size to be amenable also for larger systems), a
representative PES (from the 10 independent TL executions) was chosen
for each of the three TL classes (i.e. TL$^{\rm HF}_2$, TL$^{\rm
  B3LYP}_2$ and TL$^{\rm MP2}_2$). \\

\noindent 
H-transfer rates obtained from \textit{counting} H-transfers and dividing by total
simulation time from the MD simulations at 300 and 500~K are reported
in Table~\ref{tab:TL_htransfer_rates}. The rates at both temperatures
agree to approximately 30~\% and correspond to $\sim 1$ (34) transfers
per ns at 300 (500)~K. It is interesting to note that the trend at
300~K (i.e. lowest rate for TL$^{\rm HF}$ followed by L$^{\rm B3LYP}$
and TL$^{\rm MP2}$) reverses at 500~K. Overall, the rates for TL$^{\rm
  B3LYP}$ and TL$^{\rm MP2}$ are more similar to one another.\\

\begin{table}[h]
\begin{tabular}{lcc}
 (1/ns) & \textbf{300~K}& \textbf{500~K} \\\toprule TL$^{\rm HF}$ &
  0.9 & 41.0 \\ TL$^{\rm B3LYP}$ & 1.1 & 32.1\\ TL$^{\rm MP2}$ & 1.4 &
  33.9\\\bottomrule
\end{tabular}\caption{H-transfer rates (in ns$^{-1}$) obtained from running MD simulations using representative HL PESs.
The \textit{NVE} simulations are initialized with random momenta drawn
from a Maxwell-Boltzmann distribution corresponding to 300 and 500~K,
respectively. Each rate is obtained from an aggregate of 250~ns
(1.5~$\mu$s in total).}\label{tab:TL_htransfer_rates}
\end{table}

\begin{figure}[h!]
\centering
\includegraphics[width=1.0\textwidth]{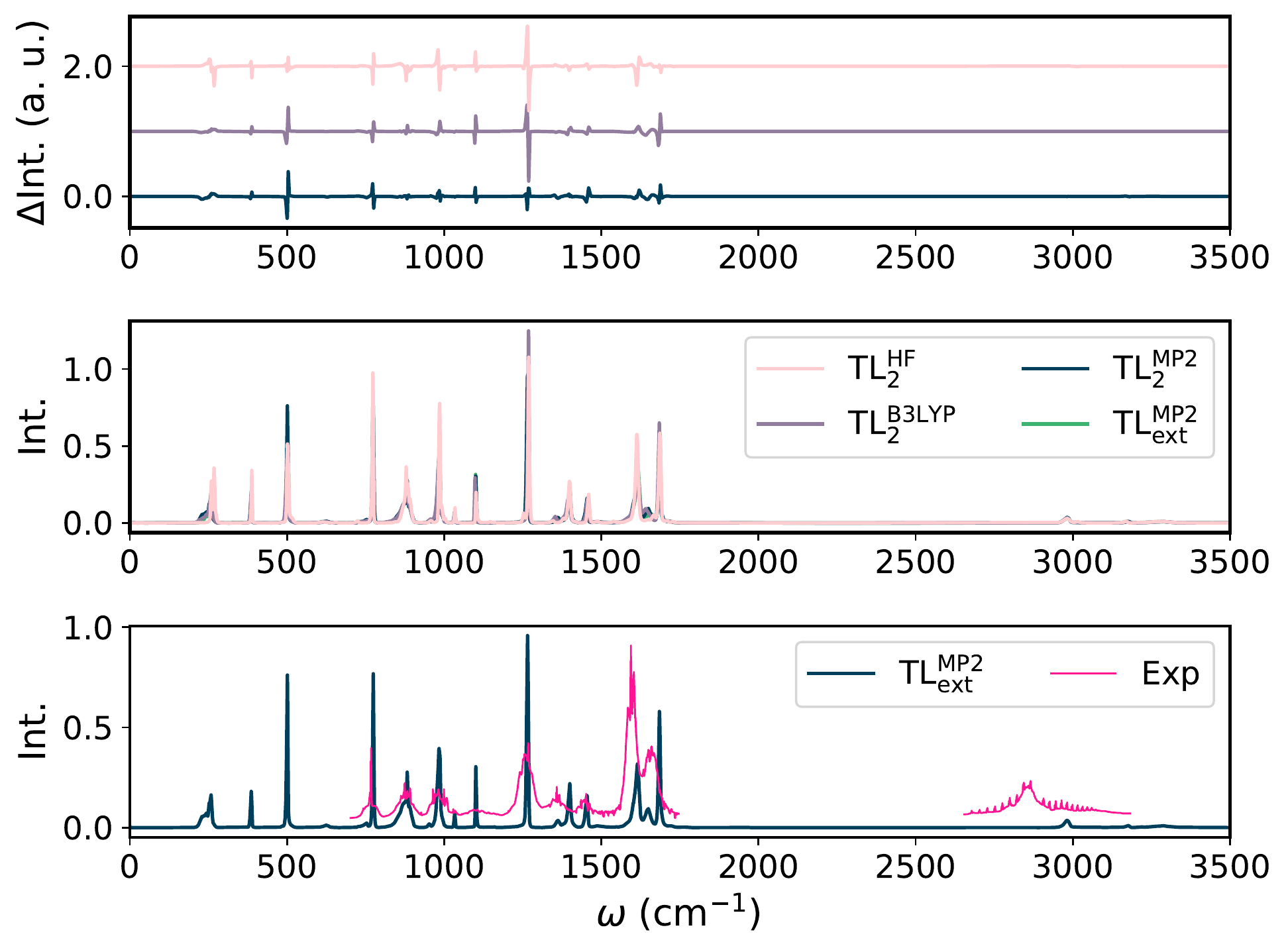}
\caption{Infrared spectra calculated from finite-$T$ MD simulations
  carried out on TL$^{\rm HF}_2$, TL$^{\rm B3LYP}_2$ and TL$^{\rm
    MP2}_2$ are shown in the middle panel. Each spectrum is averaged
  over 1000 independent trajectories of 250~ps and run with random
  momenta drawn from a Maxwell-Boltzmann distribution corresponding to
  300~K. As comparison, an additional spectrum is determined from
  TL$^{\rm MP2}_{\rm ext}$ that has been trained on the extended data
  set containing 862 HL CCSD(T) points and serves as reference. All
  computed spectra are consistent among themselves and agree
  reasonably well. The top panel shows the difference spectra of
  TL$^{\rm HF}_2$, TL$^{\rm B3LYP}_2$ and TL$^{\rm MP2}_2$ with
  respect to TL$^{\rm MP2}_{\rm ext}$ which are shifted for clarity.
  The smallest overall differences are found for TL$^{\rm MP2}$. The
  bottom panel shows the direct comparison of TL$^{\rm MP2}_2$ with
  the experimental spectrum determined in
  Reference~\citenum{luttschwager2013vibrational} and illustrate the
  excellent agreement for frequencies between 500 and
  1500~cm$^{-1}$. Somewhat larger deviations are found for peaks at
  $\sim 1650$ and 2900~cm$^{-1}$. Note that HCl impurities are
  observed at frequencies around 3000~cm$^{-1}$, see
  Reference~\citenum{luttschwager2013vibrational} for
  details.}
\label{fig:TL_ir_spectra}
\end{figure}

\noindent
The simulation run at 300~K were further used to compute IR spectra as
calculated from the dipole-dipole autocorrelation function, see
Figure~\ref{fig:TL_ir_spectra} (middle panel) for TL$^{\rm HF}_2$,
TL$^{\rm B3LYP}_2$ and TL$^{\rm MP2}_2$. As additional reference, an
IR spectrum was also calculated on TL$^{\rm MP2}_{\rm ext}$ which,
starting from the MP2 LL PES, has been trained on an extended TL data
set containing 862 CCSD(T) data points and is thus expected to yield
results closest to simulations on a -hypothetical - PhysNet PES
trained on CCSD(T)/aVTZ reference data. The spectra derived from the
dynamics on the HL PESs are very similar with one another. Single
deviations can, e.g. be found at $\sim 230$~cm$^{-1}$ or at $\sim
1650$~cm$^{-1}$, where TL$^{\rm HF}_2$ has no or only a small fraction
of the intensity in comparison to the other spectra, respectively. The
difference spectra $\Delta I(\nu) = I_{TL^{\rm MP2}_{\rm ext}}(\nu) -
I_{TL^{\rm k}_{\rm 2}}(\nu)$ (for $k$ = \{HF, B3LYP, MP2\}) shown in
the top panel of Figure ~\ref{fig:TL_ir_spectra} illustrate
differences between TL$^{\rm
  HF}_2$, TL$^{\rm B3LYP}_2$ and TL$^{\rm MP2}_2$ on the one hand and
TL$^{\rm MP2}_{\rm ext}$ as the reference. The smallest deviations are
found for TL$^{\rm MP2}_2$, as expected, while larger
differences are found for TL$^{\rm HF}$ and TL$^{\rm B3LYP}$.
For example, the intensity of the peak at $\sim 1620$~cm$^{-1}$ is twice
as large for TL$^{\rm HF}_2$ and the line positions of TL$^{\rm HF}_2$ and
TL$^{\rm B3LYP}_2$ at $\sim 1260$~cm$^{-1}$ are shifted to the blue by $\sim 3$~cm$^{-1}$ compared
to TL$^{\rm MP2}_2$ and TL$^{\rm MP2}_{\rm ext}$, respectively. The bottom panel shows the direct
comparison of TL$^{\rm MP2}$ with the experimental spectrum determined
in Reference~\citenum{luttschwager2013vibrational} and illustrates the
excellent agreement for frequencies between 500 and
1500~cm$^{-1}$. Somewhat larger deviations are found for peaks at
$\sim 1650$ and 2900~cm$^{-1}$. Aside from obtaining correct peak
positions from the calculated IR spectra, the infrared intensities are
also found to remarkably well capture the trend visible in the
experiment.

\section{Discussion}
\begin{figure}[h!]
\centering
\includegraphics[width=1.0\textwidth]{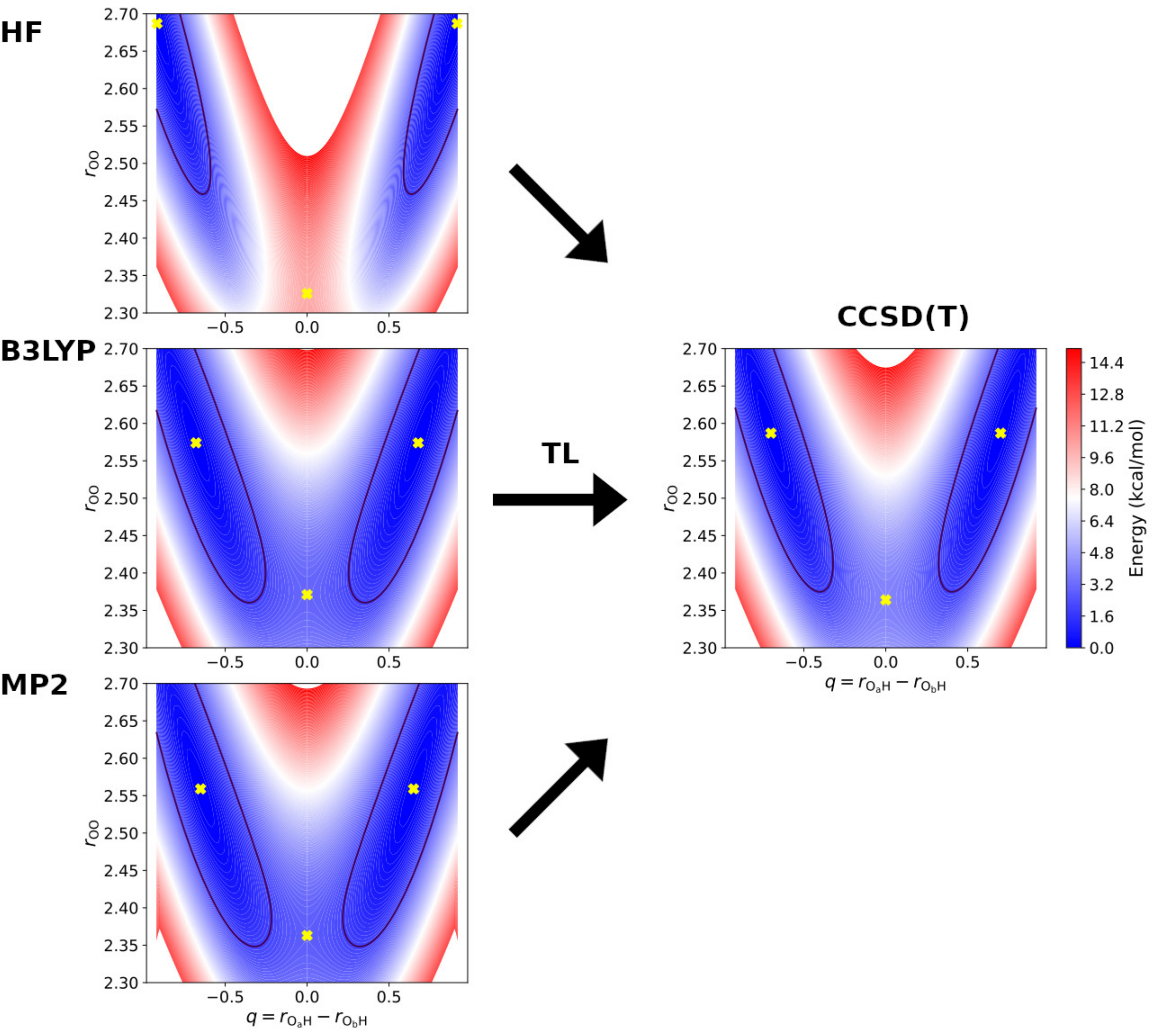}
\caption{2D representation of the LL (HF, B3LYP and MP2) and HL (TL
  with CCSD(T) target) PES spanned by the O--O distance and the
  reaction coordinate $q = r_{\rm O_AH} - r_{\rm O_BH}$. Isocontours
  of 2.0~kcal/mol are shown and the stationary points are marked by a
  gray cross.}
\label{fig:2D_representations}
\end{figure}

The present study considered TL from a range of LL methods to CCSD(T)
using variable amounts of HL information. For the molecule considered
here - malonaldehyde - the relative computational cost of the LL
methods ranged from 1 to $\sim 10^3$ but the qualities of the
transfer-learned HL PESs - as inferred from the observables considered
- were almost identical if 50 to 100 HL data points were used in the
TL step. Two-dimensional projections of the LL PESs compared with the
best HL target PES are reported in Figure
\ref{fig:2D_representations}. This provides a low-dimensional
impression of the amount of reshaping that is conveyed by the TL
step because all TL-PESs from the three LL reference
calculations yield comparable final PESs.\\

\noindent
Properties for which the performance of the transfer learned PESs were
evaluated and compared included the barrier for H-transfer, the
tunneling splittings, harmonic frequencies and IR spectra from
finite-temperature MD simulations in the gas phase. All results
suggest that the level of theory of the LL PES has only a minor
influence on the quality of the resulting HL PES, although starting
from a "better LL PES" (e.g. B3LYP/aVTZ) might have slight advantages
such as a shorter training time for TL. As an example, TL from
PhysNet$^{\rm HF}$ to CCSD(T)/aVTZ based on 100 points required
one order of magnitude more epochs than PhysNet$^{\rm B3LYP}$ or
PhysNet$^{\rm MP2}$. Thus, although transfer learning a surrogate model at the HF/aVDZ
level yields a HL model of almost identical quality compared to 
TL$^{\rm B3LYP}_2$ and TL$^{\rm MP2}_2$, the B3LYP/aVTZ
level of theory may offer the best cost/accuracy ratio.\\

\noindent
Starting from "equivalent" LL PESs in terms of representation error, a
HL PESs of gold standard CCSD(T) quality was sought after and
generated using TL. As an example for the performance of the TL
step, normal mode frequencies are considered. The three LL
PES achieve outstanding accuracy with respect to the frequencies from
\textit{ab initio} normal mode calculations at the respective levels
of theory for both, the minimum energy structure and for the
H-transfer transition state. PhysNet$^{\rm B3LYP}$ is most accurate
with MAE($\omega$) = 2.6, RMSE$(\omega) = 3.5$ cm$^{-1}$ and a maximum
absolute error of $\sim 8.7$~cm$^{-1}$ for the global minimum of
MA. This compares favourably with state of the art art PESs for
molecules of similar sizes. A recent permutationally invariant
polynomial (PIP) PES for tropolone yields MAE($\omega$)$^{\rm min}$
between 1.7 and 12.7 cm$^{-1}$ the reported MAE($\omega$)$^{\rm TS}$
ranges from 8.4 cm$^{-1}$ to 18.6 cm$^{-1}$, depending on chosen PIP
basis and other method-specific
parameters.\cite{houston2020permutationally}. PIP- and PhysNet-based
PESs for the formic acid dimer find MAE($\omega$)$^{\rm min}$ of 14.9
and 6.4~cm$^{-1}$,
respectively.\cite{qu2016ab,mm.anharmonic:2021,kaser2022fad} Comparing
the harmonic frequencies from the LL PESs (HF, B3LYP and MP2) with the
target level of theory, CCSD(T), reflects the amount of reshaping of
the PES that is required in the TL step to reach the reported
performance. This needs to be compared with differences in the
harmonic frequencies at the \textit{ab initio} HF/B3LYP/MP2 levels
compared with the CCSD(T) frequencies\cite{mm.tlrpimalonaldehyde:2022}
with an RMSE of 177, 28 and 21~cm$^{-1}$ for HF, B3LYP and MP2,
respectively. TL with only 100 HL data points correctly reshaped the
LL PESs with a MAE$(\omega) <3$ and RMSE$(\omega) <4$~cm$^{-1}$
(i.e. the same accuracy as for the most accurate LL PES, PhysNet$^{\rm
  B3LYP}$, trained on several ten thousand structures), respectively,
and impressively illustrates the data efficiency of TL. This translates
into improvements of 1 to 2 orders of magnitude in accuracy for normal
mode frequencies between reference and transfer-learned models\\

\noindent
It is also valuable to consider the accuracy of the computed harmonic
frequencies in light of the computational effort to obtain such
frequencies (especially at the CCSD(T) level of theory) based on
standard \textit{ab initio} techniques. As is well established in the
community, optimizations and frequency calculations at the CCSD(T)
level of theory are tremendously time and resource intensive,
especially for larger and complex systems. 
Starting from, e.g. PhysNet$^{\rm B3LYP}$, required only 100 HL
CCSD(T) data points (that include energies, forces and dipole moments)
to achieve close to spectroscopic accuracy (i.e. more than 60~\% of
the frequencies with sub-1~cm$^{-1}$ accuracy and a maximum deviation
of $\sim 4$~cm$^{-1}$ with respect to \textit{ab initio} CCSD(T)
reference). ML methods, especially in combination with data-efficient
TL or related $\Delta$-ML approaches, therefore offer an attractive
time-saving alternative as has been demonstrated here. This is
highlighted by recent \textit{ab initio} LCCSD(T)-F12/cc-pVTZ-F12
calculations for 15-atom tropolone, for which the optimization and
frequency calculation took 73 days on 12
cores.\cite{qu2021breaking}\\

\noindent
The present findings underscore the importance for a sufficiently
detailed LL model when aiming to construct a high-quality, versatile
PES based on as little HL data as possible. This is shown by
comparing to a model trained on the TL$_2$ (100 CCSD(T) data points)
data set from scratch, i.e. without pre-training. The resulting model is
predicts the energy barrier moderately well (a deviation of
$\sim 0.05$~kcal/mol with respect to reference), but has significantly 
larger errors for harmonic frequencies (i.e. MAE($\omega$) $> 25$~cm$^{-1}$) than all TL${\rm x}_{\rm y}$ models. In addition, starting 
MA structure optimization close to the stationary point is
necessary for meaningful results. As anticipated, the attempt to run a MD
simulation on such a PES fails entirely after a few time steps even at
a rather low temperature of 300~K, which renders the PES unsuitable
for dynamics studies.\\

\noindent
Contrary to the model trained from scratch, TL enables to obtain a
very HL representation of a global PES with very limited HL data
points and, thus, allows to carry out dynamical assessments of a
system. In this contribution this has been illustrated for the IR
spectrum of MA and compared to experiment, for which good agreement is
found. Since the TL data set contains only few structures which are
chosen rather close to the global minimum and H-transfer transition state, it
is conceivable that elevated temperatures lead to sampling structures
with stronger distortions which are not fully covered by the TL data
set. To the best of our knowledge, it is yet to be examined how a LL
PES is transformed in regions where no TL data is supplied and
presents a promising avenue for further investigation. Nevertheless,
this very attractive route employing TL to obtain HL PESs allows
CCSD(T)-quality MD simulations on the microsecond time scale for
medium sized molecules. As an indication for the reduction in
  computing time achieved for MA it is noted that 1 ns of ML/MD
  simulation takes 12 CPU hours whereas the same simulation with {\it
    ab initio} MD simulations at the CCSD(T)/aVTZ level would take $2.2 \cdot 10^7$
  hours, i.e. $\sim 2 \cdot 10^{6}$ longer, i.e. they are unfeasible. Of course such comparisons depend
  somewhat on the computer architecture and implementation used, but
  serve as an illustration that approaches as those discussed in the
  present work open possibilities to combine accuracy, system size and
  simulation lengths to approache CCSD(T)-level quality in sub$\mu$s simulations that were unachievable with more established
  methods. With regards to PES generation, TL
alleviates the problem of requiring thousands to tens of thousands of
HL \textit{ab initio} data points. This is a major step forward making
highest levels of theory accessible. The adaptation of this approach
to ML/MM simulations is a promising avenue towards stable, energy
conserving, long-time quantitative condensed phase simulations.\\

\section{Conclusion}
The present work demonstrated that even with HF/aVDZ as the LL model
high-quality full dimensional reactive PESs at the CCSD(T)/aVTZ level
can be obtained through TL. Given a shape of the PES
from a very economical level of theory, TL reshapes the
LL PES to a HL PES from which observables agree favourable with both,
experiments and models directly trained at the HL. Because the
computational effort to evaluate energies and forces of a PES based on
a NN does not depend on the level of {\it ab initio} theory used to
conceive the representation, extensive and routine sub-$\mu$s gas-phase
ML//MD and condensed-phase ML/MM//MD simulations with the solvent
treated with molecular mechanics become possible and are in reach.\\

\section*{Data Availability Statement}
The MP2/aVTZ data set taken from previous work\cite{mm.ht:2020} is available
on zenodo (\url{https://zenodo.org/record/3629239#.ZBhlJ47MJH4}) and the PhysNet
codes are available at \url{https://github.com/MMunibas/PhysNet}. Additional 
data sets generated during and/or analysed during the current study are available
from the authors upon reasonable request.

\section*{Acknowledgments}
The authors gratefully acknowledge partial financial support from the
Swiss National Science Foundation through grant $200020\_188724$, the
NCCR-MUST, and from the University of Basel.\\

\begin{figure}[h!]
\centering
\includegraphics[width=0.8\textwidth]{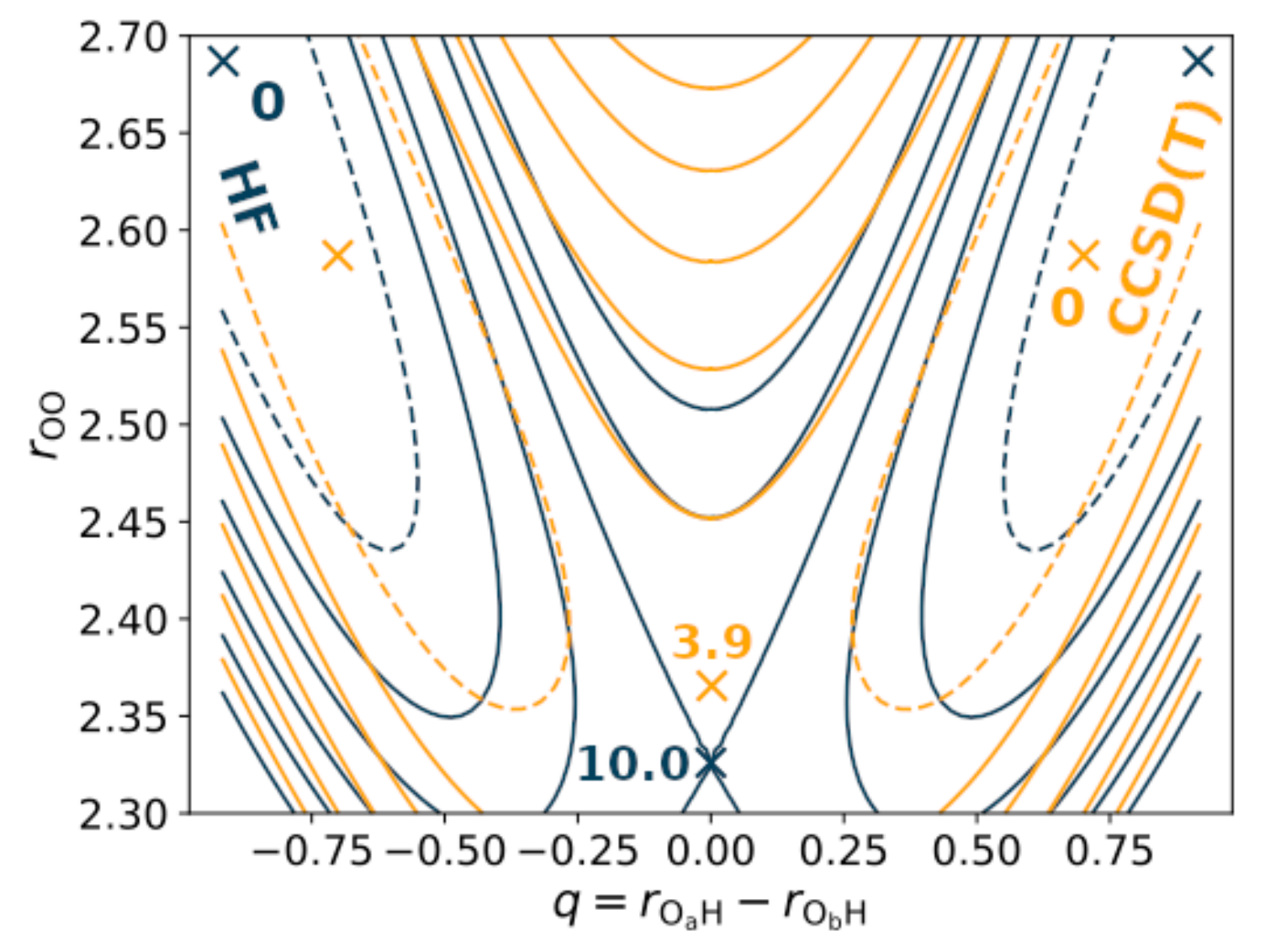}
\caption{Table of contents graphic. Comparison of the LL HF/aVDZ (blue) and HL CCSD(T)/aVTZ (orange) PES with the dashed lines indicating the 2.5 kcal/mol isocontour.}
\label{fig:toc}
\end{figure}

\clearpage
\bibliography{references}

\setcounter{figure}{0}
\setcounter{table}{0}
\renewcommand{\thepage}{S\arabic{page}}
\renewcommand{\thetable}{S\arabic{table}}
\renewcommand{\thefigure}{S\arabic{figure}}

\clearpage
\section*{Supporting Information: Transfer-Learned Potential Energy
  Surfaces: Towards Microsecond -Scale Molecular Dynamics
  Simulations in the Gas Phase at CCSD(T) Quality}

\date{\today}

\begin{table}[h]
\begin{tabular}{ccccccc}\toprule
\textbf{Mode} &  \textbf{PhysNet$^{\rm HF}$} & \textbf{HF} &  \textbf{PhysNet$^{\rm B3LYP}$} & \textbf{B3LYP} & \textbf{PhysNet$^{\rm MP2}$} & \textbf{MP2}\\
1	&	260.2	&	258.1	&	288.9	&	280.2	&	277.8	&	277.5	\\
2	&	277.9	&	267.9	&	289.9	&	287.7	&	294.7	&	286.6	\\
3	&	433.8	&	424.9	&	400.9	&	402.2	&	393.8	&	394.3	\\
4	&	529.9	&	529.2	&	521.2	&	521.6	&	512.7	&	514.1	\\
5	&	820.0	&	823.6	&	789.3	&	789.6	&	787.9	&	789.4	\\
6	&	853.9	&	858.2	&	896.1	&	894.6	&	888.7	&	888.6	\\
7	&	958.9	&	957.9	&	941.8	&	944.3	&	937.6	&	937.6	\\
8	&	1029.8	&	1032.5	&	1002.8	&	1002.1	&	1016.9	&	1012.3	\\
9	&	1118.3	&	1119.5	&	1025.8	&	1024.5	&	1025.0	&	1023.8	\\
10	&	1162.9	&	1161.9	&	1048.4	&	1048.8	&	1049.8	&	1048.8	\\
11	&	1207.7	&	1201.2	&	1120.0	&	1116.6	&	1118.0	&	1109.7	\\
12	&	1352.4	&	1369.1	&	1279.5	&	1286.8	&	1289.5	&	1288.3	\\
13	&	1505.8	&	1503.3	&	1389.8	&	1388.0	&	1405.1	&	1403.1	\\
14	&	1522.9	&	1519.6	&	1405.2	&	1405.0	&	1413.5	&	1408.0	\\
15	&	1578.2	&	1585.4	&	1480.9	&	1478.9	&	1478.6	&	1482.1	\\
16	&	1786.8	&	1776.6	&	1627.7	&	1624.4	&	1640.4	&	1641.5	\\
17	&	1930.7	&	1930.6	&	1696.2	&	1692.1	&	1692.8	&	1692.9	\\
18	&	3158.3	&	3163.2	&	2968.9	&	2966.7	&	3037.1	&	3039.0	\\
19	&	3354.5	&	3359.1	&	3106.7	&	3101.1	&	3114.4	&	3107.1	\\
20	&	3384.8	&	3387.3	&	3175.9	&	3171.3	&	3216.5	&	3217.9	\\
21	&	3896.8	&	3890.6	&	3212.8	&	3214.0	&	3269.1	&	3267.3	\\\midrule
\textbf{MAE}	&	4.8	&		&	2.6	&		&	2.5	&		\\
\textbf{RMSE}	&	6.2	&		&	3.5	&		&	3.6	&		\\\bottomrule
\end{tabular}
\caption{Harmonic frequencies of MA determined from PhysNet trained on
  different level of theory data and their comparison to the
  respective reference. Frequencies are given in cm$^{-1}$. The mean
  average and root mean squared (MAE and RMSE) errors between the
  frequencies determined from the NN and direct {\it ab initio}
  calculations at the respective level of theory are given in the last
  two lines. A corresponding table for the frequencies of the
  H-transfer TS is given in
  Tab.~\ref{sitab:harmonic_frequencies_TS}.}\label{sitab:harmonic_frequencies_opt}
\end{table}

\begin{table}[h]
\begin{tabular}{ccccccc}\toprule
\textbf{Mode} &  \textbf{PhysNet$^{\rm HF}$} & \textbf{HF} &  \textbf{PhysNet$^{\rm B3LYP}$} & \textbf{B3LYP} & \textbf{PhysNet$^{\rm MP2}$} & \textbf{MP2}\\
i	&	1761.0	&	1764.2	&	1186.0	&	1177.8	&	1193.3	&	1167.7	\\
2	&	396.2	&	394.9	&	369.4	&	369.8	&	373.7	&	373.0	\\
3	&	419.0	&	414.3	&	398.4	&	397.1	&	379.3	&	377.2	\\
4	&	619.4	&	618.2	&	576.5	&	576.0	&	562.4	&	567.3	\\
5	&	653.2	&	675.7	&	616.7	&	617.9	&	618.2	&	622.9	\\
6	&	817.3	&	812.0	&	783.8	&	785.1	&	786.4	&	787.7	\\
7	&	1025.8	&	1027.7	&	947.5	&	948.1	&	937.9	&	943.0	\\
8	&	1125.8	&	1126.8	&	1001.0	&	1002.4	&	994.0	&	992.3	\\
9	&	1141.1	&	1136.2	&	1049.8	&	1048.9	&	1051.6	&	1052.2	\\
10	&	1178.4	&	1181.6	&	1059.0	&	1060.4	&	1072.9	&	1072.6	\\
11	&	1198.7	&	1185.0	&	1110.9	&	1112.4	&	1113.1	&	1106.8	\\
12	&	1412.2	&	1406.1	&	1293.3	&	1295.7	&	1304.0	&	1301.7	\\
13	&	1417.0	&	1415.5	&	1329.9	&	1332.9	&	1340.3	&	1341.4	\\
14	&	1480.5	&	1483.9	&	1370.3	&	1371.5	&	1381.4	&	1378.9	\\
15	&	1611.0	&	1615.9	&	1507.0	&	1505.6	&	1505.5	&	1510.0	\\
16	&	1736.7	&	1737.6	&	1608.3	&	1611.4	&	1646.2	&	1646.8	\\
17	&	1819.0	&	1813.5	&	1633.6	&	1633.6	&	1666.6	&	1669.6	\\
18	&	2057.1	&	2046.2	&	1891.2	&	1879.3	&	1890.0	&	1866.4	\\
19	&	3284.7	&	3280.6	&	3080.3	&	3080.1	&	3137.4	&	3140.1	\\
20	&	3287.8	&	3281.2	&	3083.1	&	3080.4	&	3141.0	&	3140.8	\\
21	&	3410.3	&	3409.8	&	3234.5	&	3230.2	&	3291.3	&	3282.2	\\\midrule
\textbf{MAE}	&	5.1	&		&	2.3	&		&	4.9	&		\\
\textbf{RMSE}	&	7.2	&		&	3.6	&		&	8.4	&		\\\bottomrule
\end{tabular}\caption{Harmonic frequencies of the H-transfer TS of MA determined from PhysNet trained on different
level of theory data and their comparison to the respective
reference. The mean average and root mean squared (MAE and RMSE)
errors between the frequencies determined from the NN and direct {\it
  ab initio} calculations at the respective level of theory are given
in the last two lines. Frequencies are given in
cm$^{-1}$. }
\label{sitab:harmonic_frequencies_TS}
\end{table}

\begin{figure}[h!]
\centering
\includegraphics[width=1.0\textwidth]{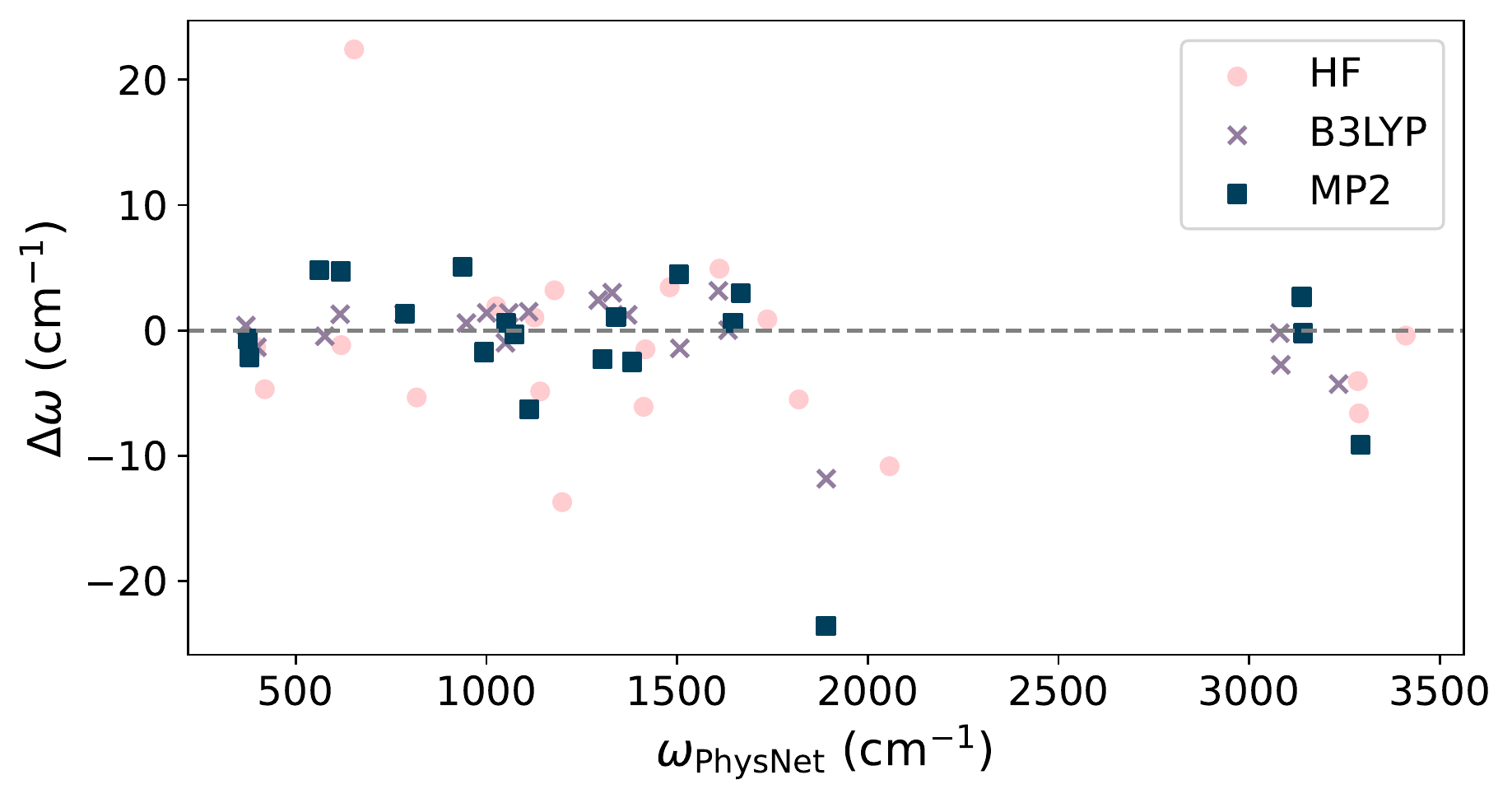}
\caption{Harmonic frequencies of the H-transfer TS of MA determined
  from PhysNet trained on different level of theory data and their
  comparison to the respective reference. Here, $\Delta\omega =
  \omega_{\rm Ref.} - \omega_{\rm PhysNet}$.}
\label{sifig:err_harmonic_frequencies_ts}
\end{figure}

\end{document}